\documentstyle[12pt]{article}

\catcode`@=12
\begin{document}

\thispagestyle{empty} 
\begin{flushright} 
DESY 98-082\\
UCRHEP-T230\\June 1998\\
\end{flushright}

\vspace{0.5in}

\begin{center}
{\Large {\bf Gauged $B - 3 L_\tau$ and Baryogenesis\\[0pt]
}}  
\vspace{1.0in}
{\bf Ernest Ma\\}
\vspace{0.1in}
{\sl Department of Physics, University of California\\}
{\sl Riverside, California 92521, USA\\}
\vspace{0.5in}
{\bf Utpal Sarkar\\}
\vspace{0.1in}
{\sl Theory Group, DESY, D-22603 Hamburg, Germany\\}
{\sl and\\}
{\sl Physical Research Laboratory, Ahmedabad 380 009, India\\}
\vspace{1.0in}
\end{center}

\begin{abstract}
It has recently been shown that by extending the minimal standard model to
include a right-handed partner to $\nu_\tau$, it is possible to gauge the $B
- 3 L_\tau$ quantum number consistently.  If we add two scalar triplets, 
one trivial ($\xi_1$) and one nontrivial ($\xi_2$) under $B - 3 L_\tau$, it 
is possible also to have desirable neutrino masses and mixing for neutrino 
oscillations.  At the same time, a lepton asymmetry can be generated in the 
early universe through the novel mechanism of the decay of the heavier $\xi_1$ 
into the lighter $\xi_2$ plus a neutral singlet ($\zeta^0$).  This lepton 
asymmetry then gets converted into a baryon asymmetry at the electroweak 
phase transition.
\end{abstract}

\newpage
\baselineskip 24pt

It has recently been shown\cite{ma} that the minimal standard 
$SU(3)_C \times SU(2)_L \times U(1)_Y$ gauge model of quarks and leptons 
may be extended to include an anomaly-free gauge factor $U(1)_{B-3L_\tau}$ 
if $\nu_\tau$ has a right-handed singlet partner, but not $\nu_e$ or 
$\nu_\mu$.  The scale of symmetry breaking of this $B-3L_\tau$ gauge group 
may even be lower\cite{maroy} than that of electroweak symmetry breaking. 
In the minimal standard model, there are dimension-six baryon-number 
nonconserving operators\cite{wein} of the form $Q^3 L$ which would induce 
the proton to decay.  In the $B-3L_\tau$ gauge model, the lowest 
dimensional baryon-number nonconserving operator is of the form $Q^9 L_\tau$, 
hence such processes are suppressed by 22 powers of some higher energy scale 
and become totally negligible.  To obtain a baryon asymmetry of 
the universe in this model, it is natural to propose instead that a 
primordial lepton asymmetry is generated\cite{fy,masark} which then gets 
converted into a baryon asymmetry during the electroweak phase 
transition\cite{ht}.

In order to generate a lepton asymmetry in the early universe, we must have 
lepton-number nonconserving interactions which also violate $C$ and $CP$, 
as well as the existence of an epoch when such processes are out of thermal 
equilibrium\cite{sakh}.  The canonical way\cite{fy} of achieving this is to 
use heavy right-handed singlet neutrinos which also allow the known 
left-handed neutrinos ($\nu_e, \nu_\mu, \nu_\tau$) to acquire small seesaw 
masses.  An equally attractive scenario was recently proposed\cite{masark} 
where heavy Higgs triplets are used.  In the $B-3L_\tau$ gauge model, since 
$\nu_e$ and $\nu_\mu$ have no right-handed singlet partners, the natural thing 
to do is to adopt a variation of the latter mechanism.  In fact, as we see 
below, the requirement of a desirable neutrino mass matrix and the absence of 
an unwanted pseudo-Goldstone boson together imply a successful leptogenesis 
scenario in this model without any further extension.

In the standard model, a general neutrino mass matrix may be obtained with 
the addition of one heavy Higgs triplet, but to generate a lepton asymmetry, 
two such triplets are required\cite{masark}.  In the minimal $B-3L_\tau$ gauge 
model, only $\nu_\tau$ gets a mass.  Furthermore, because $B-3L_\tau$ is a 
gauge symmetry, it is not obvious {\it a priori} how that will affect the 
conversion of a primordial $L_e + L_\mu$ asymmetry into a baryon asymmetry 
during the electroweak phase transition.  In this paper we will address 
both issues, {\it i.e.} neutrino masses and baryogenesis, and show how 
they may have a common solution.

The fermion content of our model is identical to that of the original 
model\cite{ma}.  The quarks and leptons transform under $SU(3)_C \times 
SU(2)_L \times U(1)_Y \times U(1)_{B-3L_\tau}$ as follows: 
\begin{equation}
\left( 
\begin{array}{c}
u_i \\ 
d_i
\end{array}
\right)_L \sim (3, 2, 1/6; 1/3), ~~~ u_{iR} \sim (3, 1, 2/3; 1/3), ~~~
d_{iR} \sim (3, 1, -1/3; 1/3);
\end{equation}
\begin{equation}
\left( 
\begin{array}{c}
\nu_e \\ 
e
\end{array}
\right)_L, \left( 
\begin{array}{c}
\nu_\mu \\ 
\mu
\end{array}
\right)_L \sim (1, 2, -1/2; 0), ~~~e_R, \mu_R \sim (1, 1, -1; 0);
\end{equation}
\begin{equation}
\left( 
\begin{array}{c}
\nu_\tau \\ 
\tau
\end{array}
\right)_L \sim (1, 2, -1/2; -3), ~~~ \tau_R \sim (1, 1, -1; -3), ~~~
\nu_{\tau R} \sim (1, 1, 0; -3).
\end{equation}
The $U(1)_{B- 3 L_{\tau}}$ gauge boson $X$ does not couple to $e$ or $\mu$ or
their corresponding neutrinos.  It can thus escape detection in most 
experiments.  Although it does couple to quarks as in a previously 
proposed model\cite{mur}, such signatures are normally overwhelmed by 
the enormous quantum-chromodynamics (QCD) background.  On the other hand, 
$X$ does couple to $\tau$ and $\nu_\tau$, so it could be observed through 
its decay into $\tau^+ \tau^-$ pairs\cite{maroy}.  Furthermore, the 
$\nu_\tau$-quark interactions in this scenario may affect the oscillations 
of neutrinos inside the sun and the earth and contribute\cite{maproy} to 
the zenith-angle dependence of the atmospheric neutrino deficit\cite{atm}. 
Since the interactions of $X$ violates $e-\mu-\tau$ universality, present 
experimental constraints limit its coupling and mass\cite{maroy}.  For 
example, $g_X < 0.1$ is required for $m_X < 50$ GeV.

The minimal scalar sector of this model consists of the standard Higgs doublet,
\begin{equation}
\left( 
\begin{array}{c}
\phi^+ \\ 
\phi^0
\end{array}
\right) \sim (1, 2, 1/2; 0)
\end{equation}
which breaks the electroweak gauge symmetry $SU(2)_L \times U(1)_Y$
down to $U(1)_{em}$ and a neutral singlet,
\begin{equation}
\chi^0 \sim (1, 1, 0; 6)
\end{equation}
which couples to $\nu_{\tau R} \nu_{\tau R}$ and breaks the $U(1)_{B-3
L_\tau}$ gauge symmetry.  The resulting theory allows $\nu_{\tau L}$ to
obtain a seesaw mass\cite{seesaw} of order 1 eV and retains $B$ as an 
additively conserved quantum number and $L_\tau$ as a multiplicatively 
conserved quantum number. 

In the original model\cite{ma}, the other two neutrinos ($\nu_e, \nu_\mu$) 
acquire masses and mix with $\nu_\tau$ radiatively.  In our present work, 
we propose instead to use the mechanism of Ref.\cite{masark} and add a 
couple of Higgs triplets:
\begin{equation}
\left( 
\begin{array}{c}
\xi_{1}^{++} \\ 
\xi_{1}^+ \\ 
\xi_{1}^0
\end{array}
\right) \sim (1, 3, 1; 0)
\hskip .3in {\rm and} \hskip .3in
\left( 
\begin{array}{c}
\xi_{2}^{++} \\ 
\xi_{2}^+ \\ 
\xi_{2}^0
\end{array}
\right) \sim (1, 3, 1; 3).
\end{equation}
Now $\xi_1$ will give small masses to $\nu_e$ and $\nu_\mu$, and will also 
generate a $\xi_2$-asymmetry of the universe when it decays at a very high 
temperature.  Furthermore, $\xi_2$ will mix $\nu_e$ and $\nu_\mu$ 
with $\nu_\tau$, and its interactions will convert the $\xi_2$-asymmetry 
into a $L_e + L_\mu$ asymmetry of the universe.  Since $B-3L_\tau$ may well 
be an unbroken gauge symmetry during the electroweak phase transition, there 
may not be any $B-3L_\tau$ asymmetry.  In that case, the total $B-L$ 
asymmetry is the same as the $-(L_e + L_\mu)$ asymmetry, which will get 
converted into the baryon asymmetry of the universe during the electroweak 
phase transition.

The above scalar sector contains a pseudo-Goldstone boson which comes
about because there are 3 global $U(1)$ symmetries in the Higgs potential
and only 2 local $U(1)$ symmetries which get broken.  In addition, there
is no $CP$-violating complex phase which is necessary for generating 
a lepton asymmetry of the universe.  However, if an extra neutral scalar 
$\zeta^0 \sim (1,1,0,-3)$ is added, then the Higgs potential will have 
additional terms which eliminate the unwanted global $U(1)$ symmetry, 
and one of these couplings will also have to be complex, allowing thus 
enough $CP$ violation to generate a lepton asymmetry, as explained below. 

The triplet scalar fields $\xi_a,~(a=1,2)$ do not acquire any vacuum 
expectation value ($vev$) to start with.  At the tree level, we can write 
down the relevant part of the Lagrangian as
\begin{eqnarray}
-{\cal L} &=& \sum_{a=1,2} M_a^2 \xi_a^\dagger \xi_a  
+ \sum_{i,j = e, \mu} f_{1ij} [\xi_1^0 \nu_i \nu_j + \xi^+_1 (\nu_i l_j 
+ l_i \nu_j)/\sqrt 2 + \xi^{++}_1 l_i l_j]   \nonumber \\&&+~ 
\sum_{i = e, \mu} f_{2i \tau} [\xi_2^0 \nu_i \nu_\tau + \xi^+_2 (\nu_i \tau 
+ l_i \nu_\tau)/\sqrt 2 + \xi^{++}_2 l_i \tau] + f_\tau \nu_{\tau R}
\nu_{\tau R} \chi^0 \nonumber \\ 
&& +~ \mu_1 [\bar \xi_1^0 \phi^0 \phi^0 + \sqrt{2} \xi^-_1
\phi^+ \phi^0 + \xi_1^{--} \phi^+ \phi^+] + \mu_2 \xi_1^\dagger \xi_2
\zeta^0 + g \zeta^0 \zeta^0 \chi^0 \nonumber \\
&& +~ h \zeta^{0*} [\bar \xi_2^0 \phi^0 \phi^0 + \sqrt{2} \xi^-_2
\phi^+ \phi^0 + \xi_2^{--} \phi^+ \phi^+] + h.c.
\end{eqnarray} 
Interactions of the scalar triplets $\xi_a,~(a=1,2)$ break the 
$L_e$ and $L_\mu$ numbers explicitly, whereas $L_\tau$ is conserved. 
Since there is no spontaneous breaking of the lepton numbers, 
there are no massless Goldstone bosons (majorons) which can contribute to the
invisible width of the $Z$ boson.  After electroweak symmetry 
breaking when the doublet acquires a nonzero $vev$, and the breaking of 
$B-3L_\tau$, there will be induced $vev$'s for these scalar triplets, 
\[
\langle \xi_1^0 \rangle \simeq {\frac{- \mu_1 v^2 }{M_1^2}} \hskip .3in
{\rm and} \hskip .3in \langle \xi_2^0 \rangle \simeq {\frac{- 
h \langle \zeta^0 \rangle v^2 }{M_2^2}} . 
\]
Since the masses of these scalar triplets and the would-be majorons are very
large, they cannot contribute to the width of the $Z$ boson.

At low energies, we can integrate out the heavier triplet fields and write 
down an effective neutrino mass matrix in the basis $\{ \bar \nu_{e L}  
~~ \bar \nu_{\mu L} ~~ \bar \nu_{\tau L} ~~ \nu_{\tau R} \}$ as,
\begin{equation}
  M_\nu = \pmatrix{ f_{1ee} \langle \xi_1^0 \rangle &f_{1e \mu} \langle 
\xi_1^0 \rangle &f_{2e \tau} \langle \xi_2^0 \rangle & 0 \cr f_{1 e \mu} 
\langle \xi_1^0 \rangle &f_{1\mu \mu} \langle \xi_1^0 \rangle & f_{2\mu \tau} 
\langle \xi_2^0 \rangle & 0 \cr f_{2 e \tau} \langle \xi_2^0 \rangle &f_{2 \mu 
\tau} \langle \xi_2^0 \rangle & 0 & m^D_\tau \cr 0 & 0 & m^D_\tau & f_\tau 
\langle \chi^0 \rangle } 
\end{equation}
where $m^D_\tau$ is the Dirac mass term for $\nu_\tau$ and $f_\tau \langle 
\chi^0 \rangle$ is the Majorana mass of $\nu_{\tau R}$.  The left-handed 
$\nu_{\tau L}$ will then get a seesaw mass, which can be of order 1 eV. 
The out-of-equilibrium condition for the generation of a lepton asymmetry 
dictates that the mass of the triplet $\xi_1$ be of order $10^{13}$ GeV, 
which implies that the $e$ and $\mu$ mass elements are also of 
order 1 eV or less.  As we will see later, $\nu_e$ and $\nu_\mu$ may 
mix with $\nu_\tau$ to form a desirable phenomenological mass matrix for 
neutrino oscillations.  The constraints on these elements come from the 
consideration of a realistic baryon asymmetry of the universe as we show 
below.

Around the time of the electroweak phase transition, we assume that 
$ B - 3 L_\tau$ is conserved; hence there cannot be any $ B - 3 L_\tau$ 
asymmetry.  In particular, $L_\tau$ is exactly conserved.  However, $L_e$ 
and $L_\mu$ numbers are broken explicitly at some high scale $M_1$ in the 
decays of the triplets $\xi_1$.  At such high energies, $SU(2)_L$ gauge 
invariance means that we need only consider one of its 
components, say $\xi_1^{++}$, which has the following decay modes:
\begin{equation}
\xi_1^{++} \rightarrow \left\{ 
\begin{array}{l@{\quad}l}
l_i^+ l_j^+ & (L_e + L_\mu = -2;~n_{\xi_2} = 0) \\ 
\xi^{++}_2 \zeta^0 & (L_e + L_\mu = -1;~n_{\xi_2} = 1) \\
\phi^+ \phi^+ & (L_e + L_\mu = 0;~n_{\xi_2} = 0) .
\end{array}
\right.
\end{equation}
Here we assumed that most of the time $\xi_2^{++} \rightarrow l_i^+ 
l_\tau^+ , $ and the other decay mode of $\xi_2$ never comes to 
equilibrium so that $L_e + L_\mu = -1$ for $\xi_2$. 
We will discuss this point later.  In the following
we will first explain how $\xi_1$ decay generates a $\xi_2$ asymmetry. 

The first decay mode does not play any role in the generation of a 
lepton asymmetry.  Here $CP$ violation comes from the interference of the 
tree-level and one-loop diagrams of Figures 1 and 2.  The scalar potential 
has one $CP$-violating phase in the product $\mu_1^* \mu_2 h$, which cannot 
be absorbed by redefinitions and produces a $\xi_2$-asymmetry when $\xi_1$ 
decays. As a result, the decays of $\xi_1^{++}$ and $\xi_1^{--}$ will create 
more $\xi_2^{++}$ 
than $\xi_2^{--}$ or vice versa; hence a ${\xi_2}$-asymmetry $\delta = 
(n_{\xi_2} - n_{\xi_2^\dagger})/n_\gamma$ will be created, given by
\begin{equation}
\delta \simeq {\frac{{Im \left[ \mu_1^* \mu_2 h
\right]} }{{16 \pi^2 g_* M_1^2}}} \left[ {\frac{{\ M_1} }{\Gamma_1}}
\right],
\end{equation}
where $g_*$ is the total number of relativistic degrees of freedom and 
\begin{equation}
\Gamma_1 = {\frac{1 }{8 \pi}} \left( {\frac{|\mu_1|^2 + |\mu_2|^2
}{M_1}} + \sum_{i,j} |f_{1ij}|^2 M_1 \right)
\end{equation}
is the 
decay rate of the triplet $\xi_1$. This $\xi_2$-asymmetry will also have 
an apparent charge asymmetry, which will be compensated by an asymmetry
in $\phi^+$ and $\phi^-$. In earlier models of leptogenesis, a lepton
asymmetry is generated when the heavy particles decay into light leptons 
and $CP$ violation enters in the vertex corrections\cite{fy} or in the 
mass matrix\cite{masark,paschos}. In contrast, we generate in the present 
scenario an asymmetry in $\xi_2$ through the quartic scalar couplings, 
which then generate a lepton asymmetry.

For the generation of the $\xi_2$-asymmetry, this decay rate should also 
satisfy the out-of-equilibrium condition\cite{kolb}
\begin{equation}
\Gamma _{1}<\sqrt{1.7g_{\ast }}{\frac{T^{2}}{M_{Pl}}}\hskip.5in{\rm at}%
~~T=M_{1},
\end{equation}
where $M_{Pl}$ is the Planck scale. We assume $M_{1} \gg M_{2}$, so that 
when $M_{1}$ decays, $\xi_2$ is essentially massless.  Taking $\mu_{1,2}/M_1 
\sim 0.1$, $f_{1ij}\sim 1$, the out-of-equilibrium condition is satisfied 
with $M_1 > 10^{14}$ GeV.  However, even if we choose $M_1 \sim 10^{13}$ GeV, 
the generated $\xi_2$-asymmetry will be only less by a factor 
$S \sim 10^{-2}$, which is still large enough to explain the baryon 
asymmetry of the universe for a value of $h \sim 10^{-4}$. This gives us 
the $e$ and $\mu$ neutrino mass matrix elements to be of order 1 eV.

At a temperature $T < M_1$, there will be a $\xi_2$-asymmetry. The decays 
of $\xi_2$ also break lepton number,
\begin{equation}
\xi_2^{++} \rightarrow \left\{ 
\begin{array}{l@{\quad}l}
l_i^+ l_\tau^+ & (L_e + L_\mu = -1) \\ 
\phi^+ \phi^+ \zeta^{0*}& (L_e + L_\mu = 0) .
\end{array}
\right. 
\end{equation}
If both of these decay modes are in equilibrium at any time, that will erase 
the lepton asymmetry of the universe\cite{ht,kolb,fy1}.  We must therefore 
require that at least one of these interactions and the scattering process, 
$$ l_i^+ l_\tau^+ \rightarrow \phi^+ \phi^+ \zeta^{0*}$$ to satisfy 
the out-of-equilibrium condition till the electroweak symmetry breaking
phase transition is over. 

For the choice $h \sim 10^{-4}$, we may take $M_2 > 10^5$ GeV to ensure
that
\begin{equation} 
\Gamma_2(\xi_2 \to \phi \phi \zeta^*) 
= {h^2 \over 16 \pi^2}{M_2 \over 8 \pi}  < \sqrt{1.7 g_*}
{T^2 \over M_{Pl}} \hskip .5in {\rm at}~ T \geq M_2
\end{equation}
so that $\xi_2$ can hardly decay into three scalars at any time. 
However, we would like the other decay mode of $\xi_2$ to be fast, 
so that the $\xi_2$-asymmetry generated during the $\xi_1$ decay gets 
converted into a lepton asymmetry.  In other words, since the number of 
$\xi_2$ is different from the numbers of $\xi_2^\dagger$, the number of 
leptons generated in decays of $\xi_2$ will be different from the number 
of antileptons generated in decays of $\xi_2^\dagger$.

We take $f_{2i} \sim 0.1$, so that the two-lepton decay
mode of $\xi_2$ is in equilibrium for most of the time,
\begin{equation}
\Gamma_2(\xi_2 \to l_i l_\tau) = {\sum_i |f_{2i \tau} |^2 \over
8 \pi } M_2 > \sqrt{1.7 g_*}
{T^2 \over M_{Pl}} \hskip .5in {\rm during} ~M_c \geq T \geq M_2
\end{equation}
where $M_c \simeq 10^9$ GeV.  During this period from $M_c$ to $M_2$, the
$\xi_2$-asymmetry will get converted into a $L_e + L_\mu$ asymmetry of the 
universe.  The interaction $\xi_2 L_i L_\tau$ will also be in 
equilibrium, which will relate their chemical potentials: 
$\mu_{\xi_2} = \mu_{L_i} + \mu_{L_\tau} $ (notations will be explained
later).  The generated $L_e + L_\mu$ asymmetry is accompanied by an equal 
amount of $L_\tau$ asymmetry.  However, that is compensated exactly by the 
$\zeta$-asymmetry created at the time of $\xi_1$ decay.  This 
$\zeta$-asymmetry generates an equal and opposite amount of $L_\tau$ asymmetry 
through $\zeta + \zeta \to \chi^* \to \nu_{\tau R} + \nu_{\tau R}$, which 
compensates the $L_\tau$ asymmetry in $\xi_2$ decay.  As a result, there 
will not be any net $L_\tau$ asymmetry as expected, since $B-3L_\tau$ 
is exactly conserved at this time.  The generated $L_{e}$ and 
$L_{\mu }$ asymmetries together give the $B-L$ asymmetry 
$$n_{L} = {1 \over 2} \delta S.$$
The above choice of parameter values will give us the neutrino
mass mixing of the $e$ and the $\mu$ to the $\tau$ neutrinos of 
order 1 eV.

We will now see how this lepton asymmetry can get converted into the 
baryon asymmetry of the universe during the electroweak phase 
transition\cite{ht}.  We consider all the particles to be 
ultrarelativistic, which is the case above the electroweak scale, but
at lower energies, although we understand that a careful analysis has to
include the mass corrections, we ignore them since 
they are small and cannot change the conclusion drastically.  The
particle asymmetry, {\it  i.e.} the difference  between the number of
particles ($n_{+}$) and the number of antiparticles ($n_{-}$) can be
given in terms of the chemical potential of the particle species $\mu$ 
(for antiparticles the chemical potential is $-\mu $) as
\begin{equation}  
n_{+}-n_{-}=n_{d}{\frac{gT^{3}}{6}}\left( {\frac{\mu}{T}}\right), 
\end{equation}  
where $n_{d}=2$ for bosons  and $n_{d}=1$ for fermions. 

In the rest of this discussion we will assume that after the triplets 
$\xi _{1}$ and $\xi_2$ have decayed, enough lepton asymmetry was 
generated.  This will give nonvanishing $\mu_{\nu e}$ and $\mu_{\nu \mu}$,  
which are directly related to $n_L$.  When these neutrinos 
interact with other particles in equilibrium, the chemical potentials 
get related by simple additive relations, and that will allow us to 
relate this lepton asymmetry $n_L$ to the baryon asymmetry during the
electroweak phase transition. 

At energies near the electroweak phase transition, most of the 
interactions are in equilibrium.  These include the sphaleron\cite{krs} 
induced electroweak $B+L$ violating interaction arising due
to the nonperturbative axial-vector anomaly\cite{thooft}.  In  Table 1, 
we give the interactions and the corresponding relations between  the 
chemical potentials.  In the third column we give the chemical potential 
which we eliminate using the given relation.  We start with chemical 
potentials of all the quarks ($\mu _{uL},\mu _{dL},\mu _{uR},\mu _{dR}$); 
the $e$ and $\mu $ leptons ($\mu _{aL},\mu _{\nu aL},\mu _{aR}$,  where 
$ a=e,\mu $); the $\tau  $ leptons ($\mu _{\tau  L},\mu  _{\nu \tau L},
\mu _{\tau R},\mu_{\nu \tau R}$); the  gauge bosons ($\mu _{W}$  for $W^{-}$, 
and 0 for all others); and the Higgs scalars ($\mu _{-}^{\phi  },
\mu _{0}^{\phi}, \mu  ^{\chi},  \mu^\zeta $).  The triplets have decayed 
away much before the electroweak phase transition and have decoupled; 
hence they do not contribute to the present analysis. 

\begin{table}[htb]
\caption {Relations among the chemical potentials}
\begin{center}
\begin{tabular}{||c|c|c||}
\hline \hline
Interactions& $\mu$ relations&$\mu $ eliminated \\
\hline
{$D_{\mu }\phi ^{\dagger}D_{\mu
}\phi $}&{$\mu _{W}=\mu _{-}^{\phi }+\mu _{0}^{\phi
}$}&{$\mu_{-}^{\phi }$}\\
{$\overline{q_{L}}\gamma _{\mu}q_{L}W^{\mu }$}&{$\mu
_{dL}=\mu _{uL}+\mu _{W}$}&{$\mu_{dL}^{{}}$}\\
{$\overline{l_{L}}\gamma _{\mu }l_{L}W^{\mu
}$}&{$\mu_{iL}^{{}}=\mu _{\nu iL}^{{}}+\mu
_{W}$}&{$\mu_{iL},i=e,\mu,\tau$}\\
{$\overline{q_{L}}u_{R}\phi ^{\dagger
}$}&{$\mu_{uR}=\mu _{0}+\mu_{uL}$}&{$\mu_{uR}^{{}}$}\\
{$\overline{q_{L}}d_{R}\phi $}&{$\mu
_{dR}=-\mu_{0}+\mu _{dL}$}&{$\mu_{dR}$}\\
{$\overline{l_{aL}}e_{aR}\phi
$}&{$\mu _{aR}^{{}}=-\mu_{0}+\mu _{aL}^{{}}$}&{$\mu_{aR},a=e,\mu$}\\
{$\overline{l_{\tau L}}e_{\tau R}\phi
$}&{$\mu _{\tau R}=-\mu_{0}+\mu _{\tau L}$}&{$\mu_{\tau R}$}\\
{$\overline{{\nu _{\tau R}}^{c}}\nu _{\tau R}\chi
$}&{$\mu ^{\chi }=- 2\mu _{\nu \tau R}$}&{$\mu ^{\chi}$}\\
{$\overline{l_{\tau L}}\nu _{\tau R}\phi
^{\dagger }$}&{$\mu_{\nu \tau R}=\mu _{0}+\mu _{\tau
L}$}&{$\mu _{\nu \tau R}$}\\
$\zeta \zeta \chi^0$&$\mu^\chi = 2 \mu^\zeta $&$\mu^\zeta$ \\
\hline \hline
\end{tabular}
\end{center}
\end{table}

We can then express all the chemical potentials in terms of the following
independent chemical potentials only,
\begin{equation}
\mu _{0}=\mu _{0}^{\phi };~~\mu _{W};~~\mu _{u}=\mu _{uL};~~\mu _{a}=\mu _{\nu
eL}=\mu _{\nu \mu L};~~\mu _{\tau }=\mu _{\nu \tau L}.
\end{equation}
We can further eliminate one of these five potentials by making use of the
relation given by the sphaleron processes.  Since the sphaleron interactions
are in equilibrium, we can write down the following $B+L$ violating relation
among the chemical potentials for three generations,
\begin{equation}
9\mu _{u}+6\mu _{W}+2\mu _{a}+\mu _{\tau }=0.
\end{equation}
We then express the baryon number, lepton numbers and the
electric charge and the hypercharge number densities in terms of these
independent chemical potentials,
\begin{eqnarray}
B &=&12\mu _{u}+6\mu _{W} \\
L_{e}=L_{\mu } &=&3\mu _{a}+2\mu _{W}-\mu _{0}  \\
L_{\tau } &=&4\mu _{\tau }+2\mu _{W}  \\
Q &=&24 \mu _{u}+(12+2m)\mu _{0}-(4+2m)\mu _{W} \\
Q_{3} &=&-(10+m)\mu _{W} 
\end{eqnarray}
where $m$ is the number of Higgs doublets $\phi$.

At temperatures above the electroweak phase transition, $T>T_{c}$, both $Q$
and $Q_{3}$ must vanish. In addition, since $B-3L_{\tau }$ is also a gauge
symmetry, this charge must also vanish. These three conditions and the
sphaleron induced $B-L$ conserving, $B+L$ violating condition can be
expressed as
\begin{eqnarray}
<Q>=0 &\Longrightarrow &\mu _{0}=\frac{-12}{6+m}\mu_u \\   
<Q_{3}>=0 &\Longrightarrow &\mu _{W}=0 \\
<B-3L_{\tau }>=0 &\Longrightarrow &\mu _{\tau }=\mu _{u} \\
{\rm Sphaleron~~ transition } &\Longrightarrow &\mu _{a}=- 5 \mu _{u}
\end{eqnarray}
Using these relations we can now write down the baryon number, lepton
number, and their combinations in terms of the $B-L$ number density, which
remains invariant under all electroweak phase transitions. They are
\begin{eqnarray}
B &=&\frac{36 + 6 m }{102 + 19 m }(B-L) \\
L_e = L_\mu &=&\frac{-78 -15 m }{204 +38m }(B-L) \\
L_\tau &=&\frac{12 + 2m }{102 + 19 m }(B-L) \\
B+L &=&\frac{-30 -7m }{102 + 19 m }(B-L)
\end{eqnarray}

We will now consider two possibilities.  In the first, the $B-3L_\tau$ gauge 
symmetry is broken after the electroweak phase transition. Then, at
temperatures below the electroweak phase transition, the other relations
remain  the same, but it is no longer neccessary to make $Q_{3}$
vanishing. Since $\phi $ acquires a $vev$, we require $\mu _{0}=0.$ With
this change we can now relate all the chemical potentials in terms of 
$\mu_u$ as
\begin{eqnarray}
<Q>=0 &\Longrightarrow &\mu _{W}=\frac{12}{2+m}\mu_u \\   
\langle \phi_0 \rangle \neq 0 &\Longrightarrow &\mu _{0}=0 \\
<B-3L_{\tau }>=0 &\Longrightarrow &\mu _{\tau }=\mu _{u} \\
{\rm Sphaleron~~ transition } &\Longrightarrow &\mu _{a}=-3 \mu_W- 5 \mu _{u}
\end{eqnarray}
This will then allow us to write down the baryon number, lepton number, and
their combinations in terms of the $B-L$ number density as
\begin{eqnarray}
B &=&\frac{48+6m}{146+19m}(B-L) \\
L_e = L_\mu &=&\frac{-114-15m}{292+38m}(B-L) \\
L_\tau &=&\frac{16+2m}{146+19m}(B-L) \\
B+L &=&\frac{-50-7m}{292+38m}(B-L) 
\end{eqnarray}

Thus after the electroweak phase transition, the $L_e + L_\mu$
asymmetry $n_L$ generated after the scalar triplets $\xi_1$ and 
then $\xi_2$ have decayed, which is equal to the $B-L$ asymmetry
at that time, will get converted into a $B$ asymmetry during the electroweak
phase transition.  Although any existing $B+L$ asymmetry gets washed out, 
we still get a non-zero $B+L$ asymmetry after the electroweak phase 
transition from the same $B-L$ asymmetry.   For consistency we check that 
the $B-L$ asymmetry remains the same during the electroweak phase transition 
and there is no $B-3L_\tau$ asymmetry.

We will now consider the other possibility when $B-3L_\tau$ is broken
before the electroweak phase transition.  In this case the electroweak
symmetry is unbroken and we still have $<Q_3>=0$, but $<B-3 L_\tau> 
\neq 0$ so that $\mu_\chi = 0$. The constraints in this case are
\begin{eqnarray}
<Q>=0 &\Longrightarrow &\mu _{0}=\frac{-12}{6+m}\mu_u \\   
<Q_{3}>=0 &\Longrightarrow &\mu _{W}=0 \\
\langle \chi \rangle \neq 0 &\Longrightarrow &\mu_{\chi}=0 \Rightarrow 
\mu_{\nu_{\tau R}} = 0;~\mu_\tau = - \mu_0 \\
{\rm Sphaleron~~ transition } &\Longrightarrow &\mu _{a}= \mu_0 - {
9 \over 2} \mu _{u}
\end{eqnarray}
The baryon and lepton asymmeties are now related by
\begin{eqnarray}
B &=&\frac{24+4m}{66+13m}(B-L) \\
L_e = L_\mu &=&\frac{-58-9m}{132+26m}(B-L) \\
L_\tau &=&\frac{16}{66+13m}(B-L) \\
B+L &=&\frac{-18-5m}{66+13m}(B-L)
\end{eqnarray}

Finally we give the relations between the baryon and lepton asymmetries
after both the electroweak symmetry and the $B-3 L_\tau$ symmetry are 
broken.  The final asymmetry will not depend on whether the electroweak
symmetry was broken before or after the $B-3 L_\tau$ symmetry is broken.
Now we have $\mu_\chi = 0$ but $<B-3 L_\tau> \neq 0$
and $\langle \phi \rangle = 0 \Rightarrow 
\mu_0 = 0 \Rightarrow \mu_\tau = 0$. We can then express the other
chemical potentials in terms of $\mu_u$ as
\begin{eqnarray}
<Q>=0 &\Longrightarrow &\mu _{W}=\frac{12}{2+m}\mu_u \\   
\langle \phi \rangle \neq 0 &\Longrightarrow &\mu _{0}= 0 \\
\langle \chi \rangle \neq 0 &\Longrightarrow &\mu_{\chi}=0 \Rightarrow 
\mu_{\nu_{\tau R}} = 0;~\mu_\tau = - \mu_0 = 0 \\
{\rm Sphaleron~~ transition } &\Longrightarrow &\mu _{a}=-3 \mu_W- {
9 \over 2} \mu _{u}
\end{eqnarray}
which then let us write the baryon and lepton numbers as some
combinations of $B-L$ as
\begin{eqnarray}
B &=&\frac{32+4m}{98+13m}(B-L) \\
L_e = L_\mu &=&\frac{-74-9m}{196+26m}(B-L) \\
L_\tau &=&\frac{8}{98+13m}(B-L) \\
B+L &=&\frac{-34-5m}{98+13m}(B-L)
\end{eqnarray}
The final baryon asymemtry of the universe is about 1/3 that of the
$B-L$ asymmetry. Hence in the present scenario, the generated $B-L$
asymmetry $n_L$ will get converted into a baryon asymmetry after the 
electroweak and $B-3L_\tau$ symmetries are broken, and the present
baryon asymmetry of the universe will be given by
\begin{equation}
n_b \sim {1 \over 6} S \delta 
\end{equation}
which is of the order of $10^{-10}$, for the choice of parameter values
we have considered earlier. 

To summarize, we studied an extension of the standard model 
to include a $B-3L_\tau$ gauge symmetry, which may be broken below the 
electroweak symmetry breaking scale. The Higgs structure is modified 
to explain the baryon asymmetry of the universe, which comes about in 
an unconventional way. We first generate an asymmetry in the number of
scalars ($n_{\xi_2}$), through only scalar interactions. This $n_{\xi_2}$
generates the $L_e + L_\mu$ asymmetry when these scalars decay. During
the electroweak phase transition, the latter gets converted
into a baryon asymmetry of the universe. Neutrino masses and mixing are
obtained naturally in this scenario. 

\vskip 0.5in
\begin{center} {ACKNOWLEDGEMENT}
\end{center}

One of us (US) would like to thank Prof W. Buchmuller and the Theory 
Division of DESY for hospitality, and to acknowledge a fellowship from the 
Alexander von Humboldt Foundation.
This work was supported in part by the U.~S.~Department 
of Energy under Grant No.~DE-FG03-94ER40837.

\baselineskip 18pt
\bibliographystyle{unsrt}

\newpage
\begin{figure}[hbt]
\vskip 2.5in\relax\noindent\hskip -.5in\relax{\includegraphics{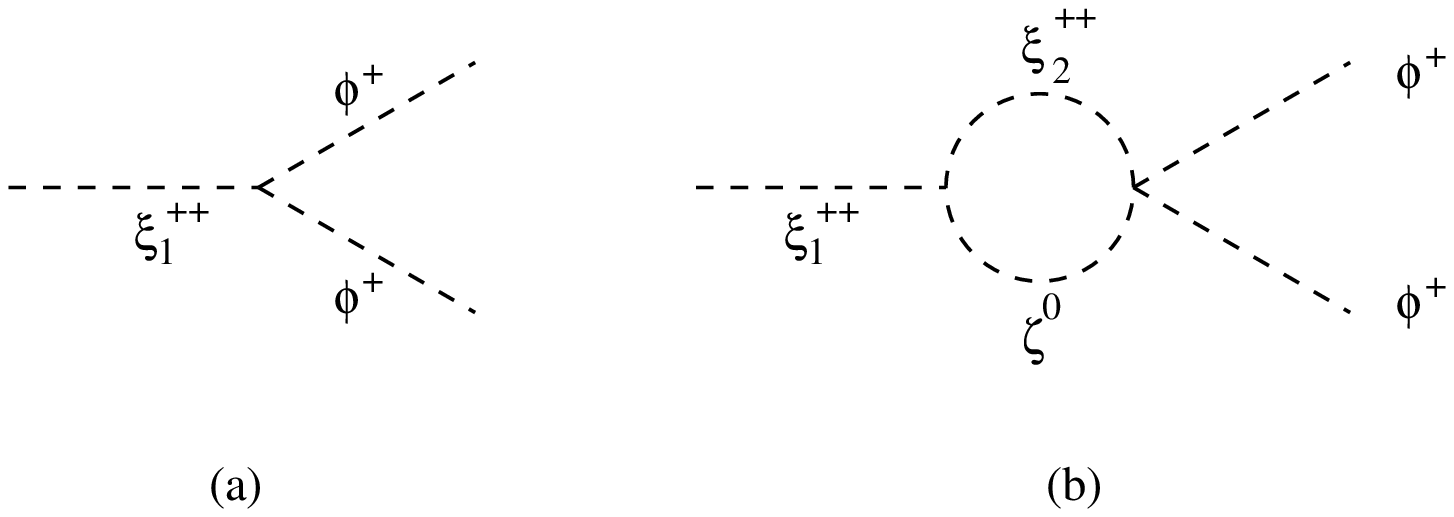}}
\caption{Tree-level and one-loop diagrams for $\xi_1 \to \phi \phi.$}
\end{figure}
\vskip 1in
\begin{figure}[hbt]
\vskip 2.5in\relax\noindent\hskip -.5in\relax{\includegraphics{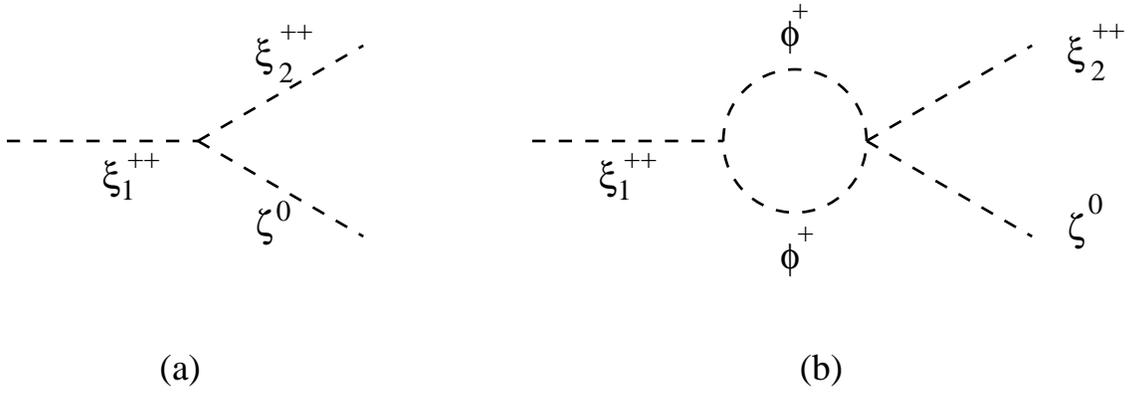}}
\caption{Tree-level and one loop diagrams for $\xi_1 \to \xi_2 \zeta.$}
\end{figure}

\end{document}